\documentclass[12pt,journal]{IEEEtran}
\usepackage{upgreek}
\usepackage{graphicx}
\usepackage{amsmath}
\usepackage{enumerate}
\usepackage{amssymb}

\usepackage[square, comma, sort&compress, numbers]{natbib}

\usepackage{caption}

\begin{document}

\title{\huge Resonant Beam Communications: Principles and Designs}

\author{ Mingliang Xiong, Qingwen Liu, Gang Wang, Georgios B. Giannakis, and Chuan Huang  


}
\maketitle

\begin{abstract}

Wireless optical communications (WOC) has carriers up to several hundred terahertz, which offers several advantages, such as ultrawide bandwidth and no electromagnetic interference.  Conventional WOC that uses light transmitters such as light emitting diodes (LEDs), directed LEDs, and lasers are facing major challenges in attenuation or tracking.  Resonant beam communications (RBCom) meets the requirements of low attenuation,  non-mechanical mobility, and multiple access. Yet, RBCom channel undergoes echo interference in contrast to common non-echo channels. An interference elimination method based on optical filtering is presented along with an exemplary interference-free RBCom system.

\end{abstract}


\IEEEpeerreviewmaketitle

\section{Introduction}
\label{sec:intr}

\IEEEPARstart{W}{ireless} optical communications (WOC) operates at several hundred terahertz, offering several advantages, such as ultrawide bandwidth and low interference~\cite{3246891,3246892,3246893}. However, conventional WOC that uses light transmitters such as light emitting diodes (LEDs), directed LEDs, and lasers are facing major challenges in path loss or mobility. In the future, the desired features of WOC are long range, high signal-to-noise ratio (SNR) transmission,  non-mechanical mobility, and multiple access~\cite{6685754,8004170}.
 
\begin{figure}
	\centering
	\includegraphics[width=3.3in]{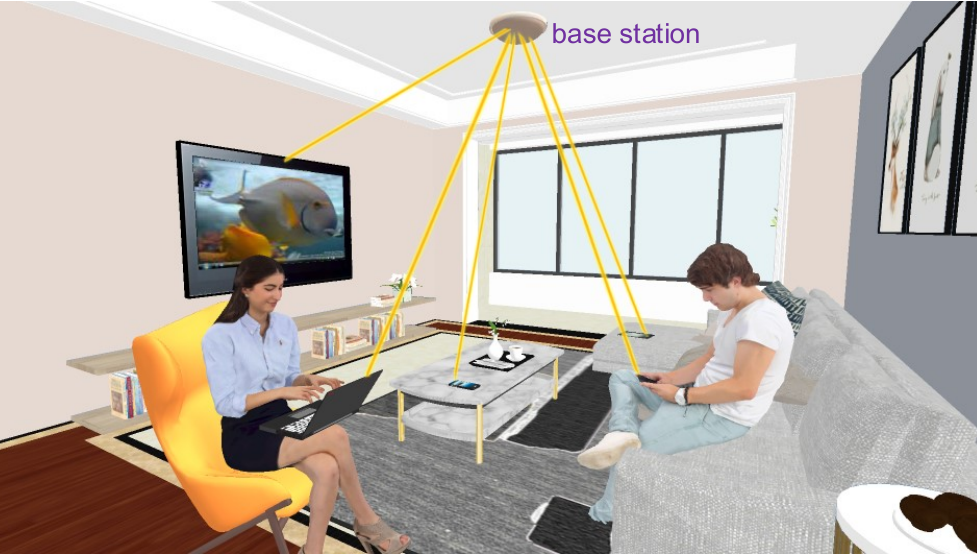}
	\caption{Resonant beam communications scenario}
	\label{fig:scenario}
\end{figure}

Resonant beam communications (RBCom) can meet those requirements. As depicted in Fig.~\ref{fig:scenario}, RBCom has the potential to  enhance mobile Internet services~\cite{a180820.09}. The so-termed resonant resonant beam system (RBS) underlies the RBCom structure. In 1973, G. J. Linford \emph{et al.} first proposed the RBS structure for fabricating a  $30~\mbox{km}$ long xenon laser and discussed its application in air pollution detection.~\cite{a190318.02}. In the same year, G.~J.~Linford \emph{et~al.} developed a $6.3$ km solid-state laser using RBS~\cite{a190318.01}.

At present, a popular application of RBS is mobile charging. As reported in ~\cite{a180727.01}, resonant beam charging (RBC), also known as distributed laser charging, has advantages such as self-alignment, intrinsic-safety, concurrent charging, high power, and no electromagnetic interference (EMI). Recent efforts have also focused on the system modeling, adaptive energy control, and charging scheduling of RBC; see e.g., \cite{a180820.07,a180820.03,qzhang2019}.

%

In 1980, a secure laser communication system based on the RBS structure was depicted in patent~\cite{US4209689A}, which highlighted the information security features provided by the RBS structure, but it did not consider the modulation speed. In high-rate RBCom, the symbol duration is shorter than the time it takes for the beam to travel from the transmitter to the receiver. In this case, echo interference emerges as a critical issue. 

The contributions of this paper are summarized as follows. Based on optical filtering, a method for eliminating the echo interference is introduced along with an exemplary design of interference-free RBCom systems. Advantages, limitations, and technical challenges of implementing a RBCom system in practice are highlighted.

\section{Wireless Optical Communications}
\label{sec:WOC}


Shannon's law suggests two ways to increase the channel capacity, namely by increasing the bandwidth or the SNR. A look back at the evolution of mobile communications, from the first generation (1G) to the fifth generation (5G), reveals that the carrier frequency increases gradually. Operating in the frequency range of infrared or visible light, WOC is desired to be utilized in the sixth generation (6G) mobile communications as it offers the following advantages.

\begin{itemize}
	\item \emph{ultrawide bandwidth}: Infrared and visible lights range over $3\mbox{--}750~\mbox{THz}$, about $10\mbox{,}000$ times of radio frequency, providing an ultrawide bandwidth.
	
	\item \emph{Low interference}: Light cannot penetrate the wall, and its direction is controllable. In addition, light is not interfered by the electromagnetic waves generated by radio systems.
\end{itemize}

\begin{figure*}
	\centering
	\includegraphics[width=5.0in]{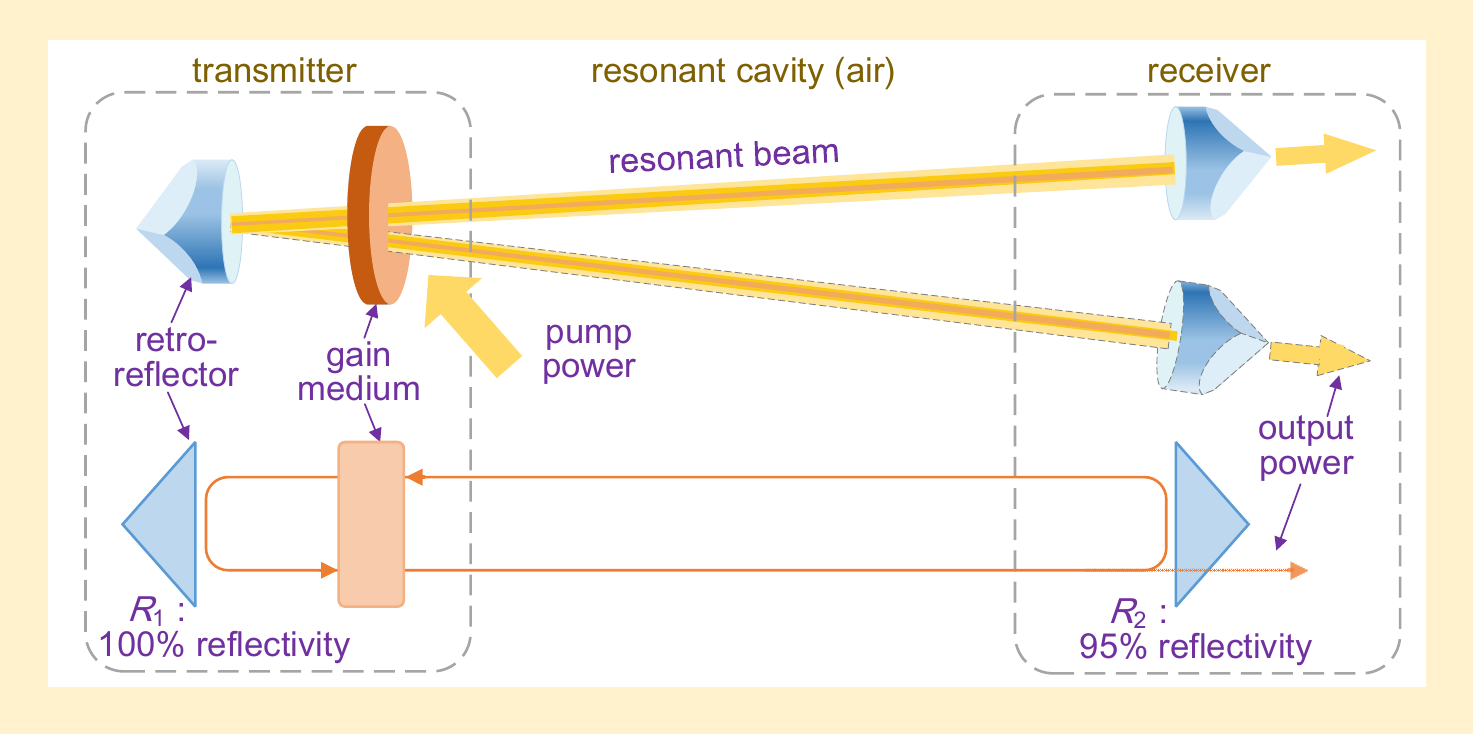}
	\caption{Resonant beam system mechanism}
	\label{fig:rbs}
\end{figure*}

\subsection{Challenges}

	WOC technologies based on  non-directed LEDs, directed LEDs, and lasers have been widely studied. Nonetheless, these technologies also face challenges.

\begin{itemize}
	\item \emph{Non-directed LED}: If adopting a non-directed LED, the base station (BS) has a wide coverage area. Each mobile station (MS) in the coverage area can receive data while moving. Besides, non-line-of-sight (NLoS) communications can be realized by receiving LED radiation  reflected by walls or ceilings. Yet, the received power as well as the SNR is low.
	
	\item \emph{Directed LED}: The light power emitted by directed LEDs is focused on the MS. Hence, the received power is larger than that generated by non-directed LEDs. A challenge that directed LEDs face is mobility. This kind of BSs must employ beam steering equipments, such as micro-electromechanical system (MEMS)-actuated mirrors, for scanning, alignment, and tracking. Nonetheless, these mechanical equipments exhibit  slow response. 
	
	\item \emph{Laser}: Lasers have advantages such as high power density, directivity, and monochromaticity.  Gratings or spatial light modulators (SLMs) can be used to change the direction of lasers by controlling the wavelength of the laser or the electrical signal applied to the SLM. These beam steering methods can offer ultra-high response. However, it is difficult to obtain the precise position of the MS.
\end{itemize}

	Conventional WOC technologies pursuit a balance between SNR and mobility. Moreover, the implementation of multiple connections is difficult when using directed LEDs or lasers.

\subsection{Requirements}
	Thinking about the future, the most significant technical requirements with respect to WOC are summarized as follows.
	
\begin{itemize}
	\item \emph{High SNR}:  Many applications such as virtual reality (VR) and augmented reality (AR) demand high-rate communication technologies.
	
	\item \emph{Mobility}: Devices such as smart phones, VR/AR, unmanned aerial vehicles (UAVs), and automobiles are expecting both communication rate and mobility. 
	
	\item \emph{Multi-access}: Generally, directional WOC technologies provide only point-to-point data transmission. In the future, multiple connections for mobile Internet of things (IoT) devices ought to be permitted.
\end{itemize}

	WOC with resonant beams can satisfy the aforementioned three requirements simultaneously. The resonant beam can be used to safely transfer several watts, while lasers are only permitted to transfer several milliwatts due to the safety constraints, which is  presented in the following section.

\section{Fundamental: Resonant Beam System}
\label{sec:rbs}

	RBS has advantages such as high power, mobility, multiple beams, and safety, which promotes its applications in wireless charging. The research on RBS structure and its mechanism lays the foundations for leveraging its potential in wireless communications.
	
\subsection{Structure}
	G. J. Linford \emph{et~al.} substituted two retroreflectors for the two reflective mirrors of the laser cavity, making it easier to create an ultralong oscillating cavity without strict alignment of retroreflectors~\cite{a190318.02}. Two important components are outlined next.

\begin{itemize}
	\item \emph{Retroreflector}: Retroreflectors, such as corner cubes and cat's eyes, reflect the incident light beam back toward the direction of the light source. An oscillating path emerges between two  retroreflectors, in which light can oscillate for many round trips until being eliminated by impacts such as absorption, partial transmission of reflectors, and geometric error. In RBS, such an oscillating path exists between the transmitter and the receiver.
	
	\item \emph{Gain medium}: The gain medium absorbs the source power and amplifies the passing beam. Some atoms at the lower energy level $E_1$ transit to the upper energy level $E_2$ while absorbing the pump power (e.g., light or electricity). In this case, if a photon passes through the gain medium, an additional photon with the same direction, frequency, and phase as the incident photon is emitted, along with the transition of an atom from $E_2$ to $E_1$.
	
\end{itemize}

	In RBS, the transmitter and the receiver, along with the air between them, constitute a laser resonant cavity, which comprises two retroreflectors (${R}_1$ and ${R}_2$), a gain medium, and a pump source, as shown in Fig.~\ref{fig:rbs}.

\subsection{Mechanism}
RBS has the capability to transfer power from the transmitter to the receiver. The resonant beam is formed by the oscillating photons between the transmitter and the receiver. If a gain medium is placed in the oscillating path (the threshold condition is achieved, that is, the gain is larger than the loss during one round trip), the power loss of the oscillating photons can be compensated. In this case, a steady state can be maintained, and the resonant beam formed by these photons will never dissipate. 

The wavelength of the resonant beam depends on the material of gain medium. For instance, if a neodymium-doped yttrium aluminum garnet (Nd:YAG) crystal is used as the gain medium, and the pump source is a $808~ \mbox{nm}$ diode laser, then $1064~ \mbox{nm}$ (approximate to $282~\mbox{THz}$) infrared resonant beam is generated.
	
In the receiver, the output power of the partial retroreflector (for example, 5 percent transmittance and 95 percent reflectivity) is proportional to the incident beam power. A photo-electric conversion device should be placed behind the retroreflector to convert the output beam to current.

The photons in the resonant cavity can be presumed to move circularly. During each round trip, a proportion of the beam power passes through the  retroreflector of the receiver, and the rest is reflected to the transmitter and amplified by the gain medium to maintain the resonance.

\subsection{Features}
	RBS features high received power, non-mechanical mobility, multiple beam links, and safety, which support the fabrication of a promising communication system.
	
\begin{figure}
	\centering
	\includegraphics[width=3.2in]{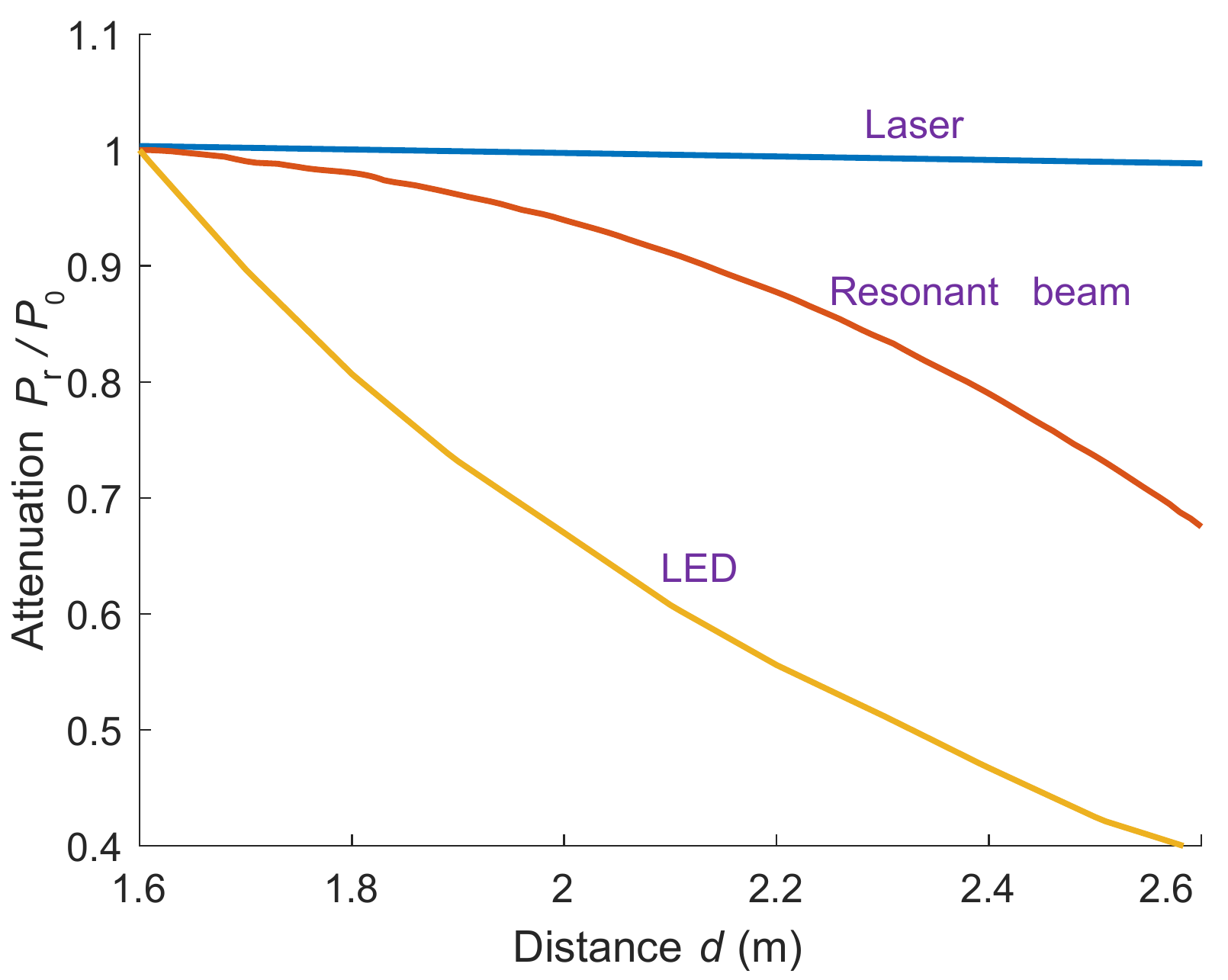}
	\caption{Path loss comparison}
	\label{fig:pathloss}
\end{figure}

\begin{itemize}
	
	\item \emph{High received power}: RBS employs a laser resonant cavity, enabling high power beam transmission. Figure~\ref{fig:pathloss} shows the path loss profiles of laser, resonant beam and LED radiation~\cite{a180820.07,a180730.01,a190606.02}. The attenuation (the ratio of the received power $P_r$ to the reference power $P_0$ at distance of $1.6~\mbox{m}$) of RBS is lower than that of LED but is higher than that of laser. These measurements of RBS come from an unoptimized experimental prototype, which can be improved in the future by optimizing the system structure, such as enhancing the manufacturing accuracy and adding a telescope.
	
	\item \emph{Non-mechanical mobility}: The self-alignment capability of RBS provides the mobility feature without the need for beam steering devices and precise positioning systems. As mentioned before, an oscillating path exists between two retroreflectors. If the threshold condition is achieved, the resonant beam forms automatically regardless of the position and the attitude angle of the retroreflectors. This mechanism allows the receiver to move while maintaining the beam resonance.
		
	\item \emph{Multiple beam links}: Between a transmitter and multiple receivers, multiple resonant beams can be generated, as long as multiple oscillation paths are available and the gain in each path is large enough. This feature supports multi-access communications.
	
	\item \emph{Safety}: Obstacles entering the cavity can break the oscillation path, preventing the beam resonance. This phenomenon allows RBS to transfer higher power to the receiver safely.
	
\end{itemize}
	
	Based on the features provided by RBS, a high-capacity, mobile, and point-to-multipoint communication system is able to be fabricated, along the lines presented in the following section.

\section{Echo Interference in Resonant Beam Communications}
\label{sec:rbcom}	

\begin{figure}
	\centering
	\includegraphics[width=3.4in]{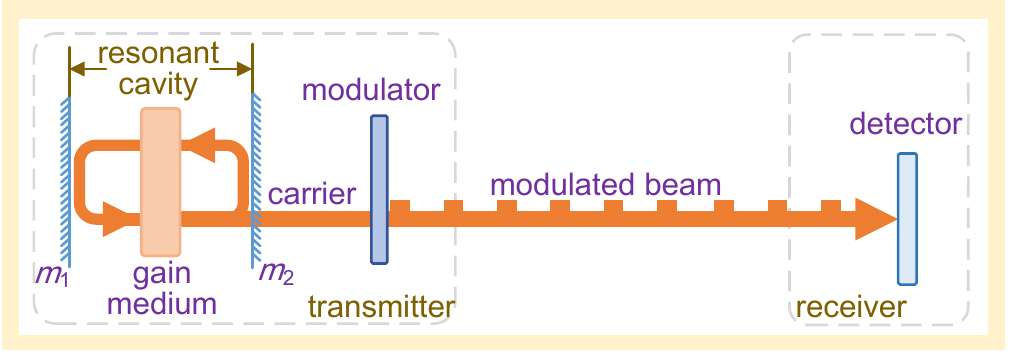}
	\caption{Laser communication system}
	\label{fig:lasercomm}
\end{figure}

\begin{figure}
	\centering
	\includegraphics[width=3.4in]{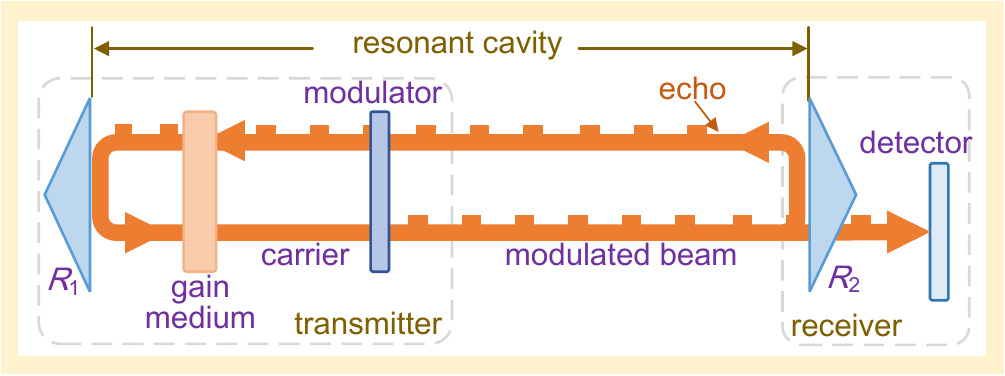}
	\caption{Echo interference in resonant beam communications system}
	\label{fig:rbcomecho}
\end{figure}

As demonstrated in Fig.~\ref{fig:rbcomecho}, RBCom adopts a resonant cavity in which  the amplitude of the carrier beam is changed by the modulator. In contrast to laser communications shown in Fig.~\ref{fig:lasercomm}, RBCom faces the echo interference issue.

\subsection{Echo Interference Issue}
In Fig.~\ref{fig:lasercomm}, the laser communication transmitter has an integrated resonant cavity, including two aligned mirrors (${m}_1$ and ${m}_2$). The laser beam is changed by the modulator and sent to the receiver. However, as shown in Fig.~\ref{fig:lasercomm}, the RBCom cavity is formed by combining the transmitter and the receiver, including two retroreflectors (${R}_1$ and ${R}_2$). Modulation is applied to the resonant beam. The receiver's retroreflector reflects the modulated beam to the transmitter, only allowing a small proportion of the beam to pass through. The reflected beam is called the echo. The following two aspects are impacted by the echo. 


\begin{figure*}
	\centering
	\includegraphics[width=6.2in]{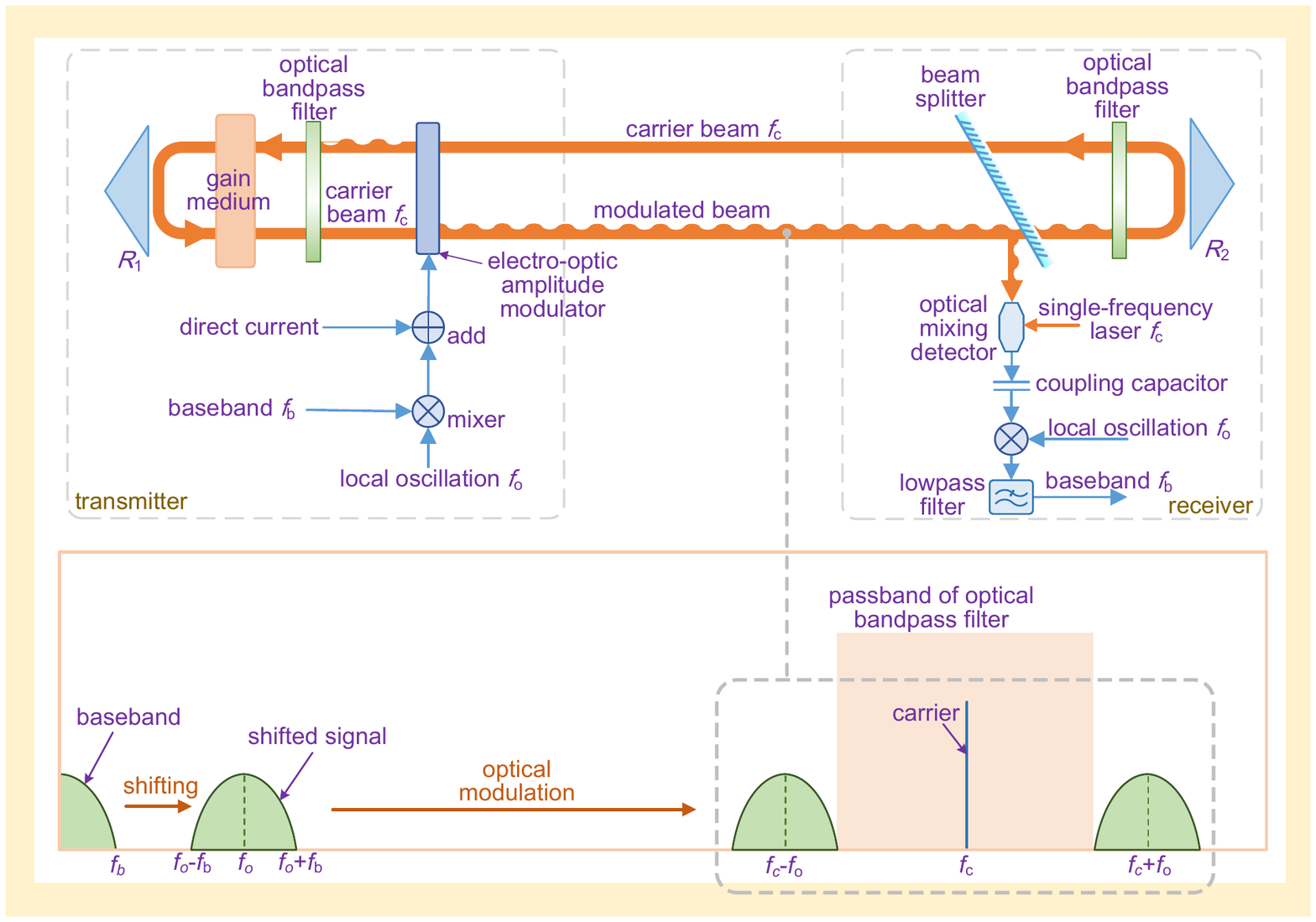}
	\caption{Exemplary design with echo interference elimination}
	\label{fig:imple}
\end{figure*} 

\begin{itemize}
	
	\item \emph{Carrier stability}: The  amplitude fluctuation of the echo impacts the density of the atoms at $E_2$ level and leads to a gain fluctuation which can break the resonating stability. Under this circumstance, the carrier amplitude can not be stable any more.
	
	\item \emph{Output signal}: The echo is amplified by the gain medium and becomes the input of the modulator. Hence, the output signal can be impacted by the echo.
	
\end{itemize}

In order to obtain a RBCom channel as good as laser communications channels, an approach to mitigate the echo effects is of paramount importance.

\subsection{Design Principle}

	The design principle is eliminating the echo interference while maintaining the beam resonance. In the following, we illustrate the reason behind this design principle and the method to achieve this.

	
\begin{itemize}
	\item\emph{Echo interference elimination}: The amplitude fluctuation of the echo is caused by band frequency components in the echo. 
	An optical bandpass filter~(OBPF) can be employed to remove these baseband components.
	
	\item\emph{Resonance maintenance}: The carrier frequency component is preserved by the OBPF to maintain the resonance. Note that the resonance emerges only when the total power loss can be fully compensated by the gain medium during one round trip.
	
\end{itemize}

	With general modulation methods, the frequency of the baseband close to $f_{\rm c}$ can not be filtered out. In order to fully filter out the baseband signal, its frequency components must be shifted out of the passband of the OBPF, as presented in the following section.

\section{Exemplary Design of Interference-Free System}
    
   	Figure~\ref{fig:imple}  depicts the interference elimination method and an exemplary system design. In this system, the baseband components are shifted out of the   passband of OBPF. Then the baseband components in the modulated beam can be filtered out easily while preserving the carrier for maintaining the resonance.
	
\subsection{Modulation}
	 Figure~\ref{fig:imple} elaborates on the three steps in modulation, that is, the baseband frequency shifting, direct-current (DC) bias, and the electro-optic amplitude modulation~(EOAM). 
	 

\begin{itemize}
	\item\emph{Baseband frequency shifting}: The baseband frequency components are shifted away from zero frequency to produce a gap between them. As depicted in Fig.~\ref{fig:imple}, the mixer is adopted to implement frequency shifting. The baseband bandwidth is $f_{\rm b}$, and the local oscillation frequency is $f_{\rm o}$.  Hence, the frequency of the shifted signal ranges around $f_{\rm o}$. Because $f_{\rm o}$ is greater than $f_{\rm b}$, the minimum frequency of the shifted signal is greater than 0.
	
	\item\emph{Direct current bias}: In Fig.~\ref{fig:imple}, the DC signal added to the shifted signal is called the bias. The transmittance of the EOAM ranges between 0 and 1, which is determined by a positive voltage that applied to the EOAM. Yet, the baseband is an alternating-current (AC) signal, which has negative periods. The DC bias raises up the  voltage of the total baseband signal to the positive region, providing an operating point, around which the transmittance of the EOAM fluctuates.	
	
	\item\emph{Electro-optic amplitude modulation}: The shifted signal is modulated on the carrier beam which is  a quasi-monochromatic wave with frequency $f_{\rm c}$. After electro-optic amplitude modulation,  the symmetric baseband components are located on both sides of the carrier frequency, leaving a gap between them. This gap is produced by the baseband frequency shifting, and it separates the baseband and the optical carrier frequency to ease the optical filtering.
	
\end{itemize}

\subsection{Demodulation}

	The optical mixing detector receives the modulated beam reflected by the beam splitter and extracts the signal carried by the beam. A coupling capacitor is employed to remove the DC component in the received signal. A mixer shifts the baseband components to the low-frequency range, and a lowpass filter removes the high-frequency components produced by this mixing process. Finally, the baseband signal can be extracted.

\subsection{Optical Filtering}

	The OBPFs are placed in the beam path to remove the baseband frequencies and preserve the optical carrier. For instance, we assume that the carrier frequency is $282~\mbox{THz}$, the baseband bandwidth  is $10~\mbox{GHz}$, and the local oscillation frequency is $20~\mbox{GHz}$. The baseband components in the modulated beam lie in two ranges, namely, from $281.97~\mbox{THz}$ to $281.99~\mbox{THz}$  and from $282.01~\mbox{THz}$ to $282.03~\mbox{THz}$. The filter passband lies in the range from $281.99~\mbox{THz}$ to $282.01~\mbox{THz}$. In this case, the filter bandwidth  $B_{\rm f}$ should be smaller than $20~\mbox{GHz}$ (i.e., the linewidth is smaller than $0.075~\mbox{nm}$).

	In Fig.~\ref{fig:imple}, there are two OBPFs in the beam path. One OBPF is mounted on the receiver to eliminate the baseband components in the echo. Another is mounted between the gain medium and the EOAM to eliminate the baseband components in the backward modulated beam generated by the modulator. 

%

\section{Conclusions and Future Work}
\label{sec:conc}

	The RBCom system operates with infrared or visible light, featuring high capacity, high SNR, non-mechanical mobility, and multi-access. The echo interference of the RBCom channel can be eliminate by optical filtering. RBCom features high SNR, non-mechanical mobility, and multiple access. It also has limitations due to its line-of-sight (LoS) feature. In the future, several challenges have to be addressed to implement RBCom in a practical system.
	
	\begin{itemize}
		\item \emph{Ultra-fast modulator}: The practical communication rate is limited by the modulating speed of the electro-optic modulator. Faster modulators are   needed to achieve higher data rates.
		
		\item \emph{Narrow-band optical filter}: There is a trade-off between the bandwidth of the optical filter and the speed of the modulator. A low-speed modulator is calling for a narrower filter bandwidth.
		
		\item \emph{Moving stability}: The motion of devices changes the length of the cavity, which disturbs the stability of resonance. In the future, to cope with motion effects are well motivated. 

		\item \emph{Duplex communication}: Uplink channel can be implemented by modulating the echo. This is important for a practical RBCom system and is expected to be studied in the future.

	\end{itemize}


%


%



\ifCLASSOPTIONcaptionsoff
  \newpage
\fi



{
\fontsize{10pt}{\baselineskip}\selectfont
\bibliographystyle{IEEEtran}
%
\bibliography{bibfile}
}

%
%

%

%


\section*{Biographies}
\fontsize{10pt}{12pt}\selectfont
\setlength{\parindent}{0em}

\textbf{Mingliang Xiong} (xiongml@tongji.edu.cn) received his B.Eng.
degree in communications engineering from the Nanjing University
of Posts and Telecommunications, China, in 2017. He
is currently a Ph.D. student in the College of Electronics and
Information Engineering, Tongji University, Shanghai, China. His
research interests include wireless optical communications and
wireless power transfer.

\vspace{6pt}

\textbf{Qingwen Liu} [M'07, SM'15] (qliu@tongji.edu.cn).) received the B.S. degree in electrical engineering and information science from the University of Science and Technology of China, Hefei, China, in 2001, and the M.S. and Ph.D. degrees from the Department of Electrical and Computer Engineering, University of Minnesota, Minneapolis, MN, USA, in 2003 and 2006, respectively. He is currently a professor with the College of Electronics and Information Engineering, Tongji University, Shanghai, China. His research interests lie in the areas of wireless power transfer and Internet of Things. 

\vspace{6pt}

\textbf{Gang Wang}
[M'18] (gangwang@umn.edu) received his B.Eng.
in electrical engineering and automation from the Beijing Institute
of Technology, China, in 2011, and his Ph.D. in electrical
engineering from the University of Minnesota in 2018, where he
is currently a postdoctoral associate. His research interests focus
on the areas of statistical signal processing and machine learning
with applications to data science and smart grids. He received
paper awards at the 2017 European Signal Processing Conference
and the 2019 IEEE Power \& Energy Society General Meeting.
He is currently on the Editorial Board of Signal Processing.

\vspace{6pt}

\textbf{Georgios B. Giannakis} [F'97] (georgios@umn.edu) received his Ph.D. degree from the University of Southern California in 1986.  Since 1999 he has been a Professor with the University of Minnesota, Minneapolis, MN, USA, where he holds an ADC Endowed Chair, a McKnight Presidential Chair in Electrical and Computer Engineering, and serves as the Director of the Digital Technology Center. His interests are in the areas of communications, networking, and statistical signal processing. He is the (co)recipient of 9 best paper awards from the IEEE Communications and Signal Processing Societies (SPS). He is a Fellow of the IEEE and EURASIP, has received technical achievement awards from the IEEE-SPS and EURASIP.
\vspace{6pt}

\textbf{Chuan Huang} [S'09, M'13] (huangch@uestc.edu.cn) is currently
with the Chinese University of Hongkong, Shenzhen. He
received his Ph.D. in electrical engineering from Texas A\&M
University in 2012, and his M.S. in communications engineering
and B.S. in mathematics, from the University of Electronic Science
and Technology of China. He was a postdoctoral research
fellow from 2012 to 2013, and an assistant research professor
from 2013 to 2014, at Arizona State University. He currently
serves as an Editor of IEEE Access and IEEE Wireless Communications
Letters. His current research interests include wireless
communications and machine learning.





\end{document}